\documentstyle[preprint,tighten,aps]{revtex}
\def\lapprox{\mathrel{\mathop
  {\hbox{\lower0.5ex\hbox{$\sim$}\kern-0.8em\lower-0.7ex\hbox{$<$}}}}}
\def\gapprox{\mathrel{\mathop
  {\hbox{\lower0.5ex\hbox{$\sim$}\kern-0.8em\lower-0.7ex\hbox{$>$}}}}}

\begin{document}
\author{  S. Cassisi $^1$, V. Castellani $^2$,
 S. Degl'Innocenti $^2$, G. Fiorentini $^3$ and
	B. Ricci $^3$}

\address{
  $^{1}$ Osservatorio Astronomico di Collurania, via Maggini10,
        I-64100 Teramo, Italy.\\
  $^{2}$ Dipartimento di Fisica dell'Universit\`a di Pisa and
        Istituto Nazionale di Fisica Nucleare, Sezione di Pisa,
	via Livornese 582/A, S. Piero a Grado, 56100 Pisa.\\
  $^{3}$ Dipartimento di Fisica dell'Universit\`a di Ferrara and
        Istituto Nazionale di Fisica Nucleare, Sezione di Ferrara,
        via Paradiso 12, I-44100 Ferrara, Italy .
}

\preprint{\vbox{\noindent
          \null\hfill  INFNFE-02-00}}

\title{Stellar Evolution and Large Extra Dimensions}

\date{February 2000}

 \maketitle

\begin{abstract}

We discuss in  detail the information on large extra dimensions
which can be derived in the framework of stellar evolution theory
and observation. 
The main effect of large extra dimensions arises from
the production  of the Kaluza-Klein (KK) excitations of
the graviton. 
The KK-graviton and matter interactions are of gravitational
strength, so the KK states never become thermalized and always 
freely escape.
In this paper we first pay attention to the sun.
Production of KK gravitons is 
incompatible with helioseismic constraints unless  
the $4+n$ dimensional Planck mass $M_s$ exceeds
$300$ Gev/c$^2$.
Next we show that stellar structures
in their advanced phase of H burning evolution put much more severe
constraints, $M_s > 3-4$  TeV/c$^2$, improving on
current laboratory lower limits. 

\end{abstract}

\section {Introduction}

Recently there has been a revived interest in the physics of
extra-spatial dimensions.
In order to provide a framework of solving the hierarchy problem,
in refs. \cite{Arkani98a,Antoniadis98,Arkani98b},  
the fundamental Planck scale -  where gravity becomes comparable
in strength with the other interactions -
was taken to be near the weak scale. The observed 
weakness of gravity at long distances  is due to the
presence of $n$ new spatial dimensions, with size $R$
which are large compared to the
electroweak scale. The relation between the Planck mass
in $4$ dimensions 
($M_{Pl}= \sqrt{\hbar c/G_N}=1.2 \, 10^{19}$GeV/c$^2$)
and that in $4+n$ dimensions ($M_s$) is
\begin{equation}
\label{eq1}
R^n= (\hbar/c)^n M_{Pl}^2 / (M_s^{n+2} \, \Omega_n)
\end{equation}
where $\Omega_n$ is the volume of the n-dimensional
sphere with unit radius.
Laboratory limits, essentially from LEP II \cite{Lep} give a lower
bound on $M_s$ of about 1 TeV/c$^2$.
The choice  $M_s \sim 1$ TeV/c$^2$  yields $R \sim 10^{32/n -17}$ cm.
The case $n=1$ gives $R\simeq 10^{15}$cm which is 
excluded since it would modify newtonian gravitation at
solar system distances. Already for $n=2$ one has $R\simeq 1$mm
which is  the distance where our present
experimental measurement of gravitational  forces stops,
and one needs information from different sources.

In this context, one should remind that in last decades the improved 
knowledge of several physical mechanisms has 
allowed  astrophysicists to produce stellar models with a significant degree of 
reliability. As a matter of  fact, current stellar  models  nicely 
reproduce the large variety of stellar structures populating the sky, passing also 
some subtle tests as the ones recently provided by seismologic investigations of 
our sun. Such a success has already opened the way of using stellar structures as a 
natural laboratory to test the space allowed for  new physics, i.e., to investigate 
the allowed modifications of the current physical scenario \cite{Raffeltlibro}. 
This looks as a 
quite relevant opportunity, bearing in mind that a stellar structure is governed 
by the whole ensemble of physical laws  investigated in terrestrial laboratories 
and that these stellar structures, in varying their mass and ages, experience a 
range of physical situations not yet reached in current laboratory experiments. 
On this basis, the ``stellar laboratory" has already provided relevant 
constraints on several physical ingredient as, e.g., the existence of Weak 
Interacting Massive Particles \cite{Rood89,SF88} or the neutrino 
magnetic moments \cite{Raffelt90,Castellani93}.

Astrophysical constraints on large extra dimensions have
been discussed in \cite{Arkani98b} and 
in \cite{Barger99}. 
The main effect of large extra dimensions arises from
the production  of the Kaluza-Klein (KK) excitations of
the graviton. The KK-graviton and matter interactions are of gravitational
strength, so the KK states never become thermalized and always 
freely escape. The associated energy loss (through photon-photon annihilation, 
electron-positron annihilation,
gravi-Compton-Primakoff scattering, gravi-bremsstrahlung,
nucleon-nucleon bremsstrahlung) 
have been calculated in \cite{Barger99} and observational constraints on $M_s$
have been derived from simple considerations on 
the energetics of sun, red giants and supernovae.

In this paper  we discuss in more detail the information on large
extra dimensions
which can be derived in the framework of stellar evolution theory
and observation. 
The first part is devoted to the study of the 
sun, which represents a privileged laboratory in view of the
richness and accuracy of available data.
In particular we shall consider  the following topics:\\
i)As well known there is a remarkable agreement
between the predictions of the Standard Solar Model (SSM) and the results
of helioseismic observations, see e.g. \cite{eliosnoi,nutel99,Bahcall99}.
Production of KK gravitons provides a new energy loss, which 
will become incompatible with helioseismic constraints if 
the $4+n$ dimensional Planck mass $M_s$ is sufficiently low.
In this way we shall determine lower limits 
on $M_s$ from helioseismic observations.\\
ii)Despite its several successes, the standard solar model presents
us with some puzzles, e.g. the  deficit of solar neutrinos, see e.g. \cite{valencia97}, 
the  depletion of the photospheric lithium abundance,
see e.g. \cite{litio}, and - perhaps - an underestimate 
of the sound speed just
below the convective envelope, see e.g. \cite{nutel99,Bahcall98}. 
Could it be that the new physics of KK-graviton 
production accounts for some of these anomalies?

In addition, the efficiency of  KK-graviton energy loss 
 appears strongly dependent on the temperature. This
suggests to consider stars
experiencing internal temperatures much larger than in the sun and which, in 
turn, are  particularly sensitive to the efficiency of cooling 
mechanisms. In this way the investigation will be extended to red giants structures, which
will provide a much more stringent limit on $M_s$.

In  section \ref{first} we give a first look at 
KK-graviton production in the sun, determining the order of magnitude of
the acceptable $M_s$ and presenting the structure of solar
models where the new energy loss is relevant.
In section \ref{helio} we shall determine the helioseismic
constraints on $M_s$, from  data on the photospheric 
helium abundance and on sound speed in the energy production
region. The effect of KK-graviton production
on the ``solar puzzles'' mentioned
above  is discussed in sect. \ref{puzzle}.
Sect. \ref{rg} will be   devoted to red giant stars.

Our conclusions are summarized at the end of the paper,
whereas in the appendix we collect the relevant formulas 
for the energy losses.

\section{A  first look at the effects of KK-graviton production in the sun}
\label{first}

It is interesting to compare the energy loss 
due to KK-graviton production with the energy production from the pp
chain at the center of the sun. The values of density, temperature 
and chemical composition derived from the SSM of \cite{Bahcall98}
are presented in Table \ref{tabssm}. The results of other SSM calculations
are similar, see e.g. \cite{report}.
Energy loss and production rates, computed according 
to the results of Appendix A, are compared in Table \ref{tabepsi}.
The most important contribution always comes from the photon-photon
annihilation.  

One expects that the solar solar structure would be drastically
modified if the energy loss due to 
KK-gravitons becomes comparable with the nuclear energy production
rate. In this way one can derive the following lower limits on $M_s$:
\begin{equation}
\label{ms2}
n=2 \,: \quad M_s> 140 \, GeV/c^2
\end{equation}
\begin{equation}
\label{ms3}
n=3 \,: \quad M_s>  3.5 \, GeV/c^2
\end{equation}
This result which is essentially the same as that in ref. \cite{Barger99}
suggests that we concentrate on the $n=2$ case only.
So far we assumed just a rough knowledge of the solar structure. 
One can expect that
more detailed information, as that provided by helioseismology, 
provide more stringent constraints.

To understand in more detail the effect of KK-graviton production
we have built solar models which include this additional
energy loss. The energy generation subroutine was modified so as to include
the KK-graviton loss and the stellar evolution code FRANEC \cite{Ciacio}
was run by varying the three free parameters of the model (initial
helium abundance $Y_{in}$, initial metal abundance $Z_{in}$
and mixing length $\alpha$) until
it provides a solar structure 
(i.e. it reproduces the observed solar luminosity, radius
and photospheric metal abundance  at the solar age).

As an example, we present here the case $M_s=0.2$ TeV/c$^2$.
The main differences with respect to our SSM are depicted
in Table \ref{tab02} and Fig. \ref{fig02}.
Several features can be easily understood by observing
that the solar model with KK-graviton production has to
produce, now and in the past, a higher amount 
of nuclear energy, in order to compensate 
for the additional energy loss.
 
More hydrogen has been burnt into helium, and 
the initial helium abundance has to be reduced with 
respect to the SSM (otherwise one would get 
a stellar structure  which, being too much helium rich,
would be presently overluminous). Consequently,
the present photospheric helium abundance $Y_{ph}$
is decreased with respect to the SSM prediction.

Nevertheless, the central helium abundance is still
somehow larger than in SSM and more energy is being
produced in order to compensate for the KK-graviton losses.
This is achieved with a somehow larger central temperature.

In the solar core, both temperature and ``mean molecular
weight'' are thus higher than in the SSM, so that one cannot
a priori decide for the behaviour of the sound speed.
In fact Fig. \ref{fig02} shows a decrease near the center and a
significant increase near $R=0.2 R_\odot$, i.e. in a region where
helioseismic determinations are still very accurate.

These observations will be useful for determining the 
relevant observables which are sensitive to $M_s$ and which
can be constrained by means of helioseismology.

In Fig. \ref{figepsi} we also compare the nuclear energy production rate
with the losses due to KK-gravitons.  The results are consistent
with the qualitative energetics analysis discussed above.

\section{Helioseismic constraints  on $ M_s$}
\label{helio}

Helioseismology  provides detailed information  on several solar properties. 
In particular, the sound speed profile and the photospheric helium 
abundance  $Y_{ph}$ are determined with high accuracy.
In  ref. \cite{eliosnoi} it was estimated that the isothermal sound
 speed squared, $u= P/\rho $ at distance
$R=0.2 R_\odot $ is determined with an accuracy of about 
                          $\Delta u/u \approx 1\cdot 10^{-3}$,
\begin{equation}
\label{equ02}
u_{0.2}^{\odot}=(1.238 \pm 0.001)\cdot 10^ {15} \, cm^2/s^2
\end{equation}
This uncertainty, defined as  the ``statistical'' \cite{eliosnoi}
 or ``one sigma'' error  \cite{valencia97},
 was obtained 
 by taking into account all possible contributions 
 arising from: i) measurement errors, ii) the inversion method
and  iii) the choice of the reference model
 (the recent analysis of \cite{Bahcall99} confirms  the estimate
 of \cite{eliosnoi} for each contribution to the  uncertainty). 
 These  estimated errors were  added in quadrature.
With a similar  attitude  the uncertainty of  
$Y_{ph}$ was estimated:
\begin{equation}
\label{eqyph}
     Y_{ph}^{\odot}=0.249\pm0.003
\end{equation}
  
Recent accurate standard solar model calculations 
are successful in reproducing  sound speed in the energy production
region as well as the photospheric helium abundance, 
their predictions being quite close to the central helioseismic
estimates, see e.g. \cite{Bahcall98,eliosnoi}. 
On the other hand, as discussed in the previous section, 
 both quantities are sensitive 
to the energy loss due to KK-graviton production 
 For this reason, we  concentrate here on 
$Y_{ph}$ and on  the value of  
$u$ at $R=0.2 R_\odot$ {hereafter $u_{0.2}$. 

We have built a series of solar models with $M_s$ 
in the range of few hundred GeV/c$^2$  in order to determine the
dependence of both observables on $M_s$, see Figs. \ref{figdelta} 
and \ref{figpiano}.  
For each observable $Q$ the results have been parametrized in the form
\begin{equation}
\label{eqfit}
          Q(M_s) =Q_{SSM} ( 1 + (m/M_s)^\alpha)\quad ,
\end{equation}
with the  results for the parameters   $ m $ and  $\alpha$ shown 
in Table \ref{tabalfa}. By requiring  that the differences
in the calculated observables do not exceed the  helioseismic uncertainty, 
we get the following lower bounds on $M_s$:\\

\noindent
-from $Y_{ph}$ :   $M_s > 0.23$ TeV/c$^2$\\

\noindent
-from $u_{0.2}$ : $M_s > 0.31$ TeV/c$^2$\\

These bounds  are stronger that that of Eq. (\ref{ms2}), which
was obtained by using crude energetical considerations. 
However, the accuracy of helioseismic method has yielded an improvement
of just a factor of two.
In fact, KK-graviton energy loss rate $\epsilon_{KK}$ depends
on high powers of $M_s$, so that drastic changements of 
 $\epsilon_{KK}$ result from just tiny modifications of $M_s$.

\section{Extra-dimensions and the puzzles of the SSM}
\label{puzzle}

As well known, in front of its several successes,
 the standard solar model presents us with some puzzles: 

\noindent
i) The signals  measured by all solar neutrinos experiments are 
systematically lower than those   predicted 
by SSMs,  an effect which is now  
commonly ascribed to
 neutrino oscillations.

\noindent
ii) The  observed photospheric lithium abundance is   
 a factor  of hundred  smaller than the  meteoritic value
\cite{litio}. 
Lithium is  being continuosly mixed in the convective 
envelope, however -according to the SSM - it should not be 
destroyed by nuclear reactions since even  at the bottom of the
 convective zone the temperature is not high enough to burn it. 
This signals some deficiency of the standard solar model, 
which is built in a one dimensional approximation and neglects rotation, 
see \cite{RVCD}.

\noindent
iii) The helioseismically determined sound speed just below 
the convective envelope is somehow smaller
(by 0.4\%) with respect to the predictions of the most 
recent and accurate SSM calculations, see e.g. \cite{nutel99,Bahcall98}.

It is thus natural to ask what is the effect of the hypothetical 
large extra dimensions on these items.
{\footnote 
{We recall that in \cite{Smirnov} conversion of electron neutrinos  to 
the light fermions propagating in the bulk of $4+n$ dimensions has been
considered as a solution of the solar neutrino problem.}}

Concerning solar neutrinos, the answer is already contained in the  
previous discussion. When KK-graviton  production is  effective,
 the central temperature increases and consequently the
 production of  Beryllium and Boron neutrinos
is increased, see Table \ref{tab02}. KK-graviton production would 
thus make the neutrino puzzle even more serious.
At the bottom of the convective zone the temperature
 would be even smaller than that predicted by SSM, see Table \ref{tab02}, 
so that there is no help in lithium burning. Sound speed
 just below the convective enevelope is practically unchanged
 with respect the SSM,
so that the disagreement cannot be affected.

In short, KK graviton production would provide no cure  to the  SSM puzzles.

\section{Red giants and KK-gravitons}
\label{rg}

A glance at the  current evolutionary scenario easily indicates low-mass Red 
Giant Branch (RGB) stars as good candidates for 
investigating the effects of KK-graviton production.
As a  matter of fact a RGB star reaches internal temperatures of the order of 
$10^8$ K. 
Moreover, the structure of RGB stars is quite sensitive
 to the cooling mechanisms which regulate the 
size of the He core at the He ignition.
The size of He core in turn  governs several 
observational quantities both in these RGB structures as well as in the 
subsequent phase of central He burning (Horizontal Branch, HB) stars. 
We will follow this approach discussing the effect of KK-graviton cooling on 
the evolution of suitable RGB structures. Comparison of theoretical predictions 
with available experimental (i.e. observational) data will allow to put more 
stringent constraints on the minimal $4+n$ dimensional  Planck mass $M_s$.

To perform our investigation we used our latest version of the FRANEC
evolutionary code \cite{Ciacio} to predict the observational
properties of stellar models with different metallicities but with a
common age of the order of 10 Gyr, thus adequate for RGB stars
actually evolving in galactic globular stellar clusters (GCs). In
order to make more clear to the reader the following discussion, in
Fig. \ref{fighr} we show the typical Hertzprung-Russel diagram for a
galactic GC (upper panel) and the corresponding theoretical one (lower
panel) as obtained by using the prescriptions provided by our own
computations. The most relevant evolutionary phases and observational
features are clearly marked. The diagram represents the locus of stars
for a given chemical composition and age but different masses. As the
mass increases, the star moves from the Main Sequence location (H
central burning phase) to the RGB (H shell burning phase) till
reaching a maximum luminosity where the central He ignition occurs
(RGB tip), driving the structure to the central He burning (Horizontal
Branch) phase.

Numerical experiments disclose that a stronger  cooling has a little effect on 
the morphology of the diagram depicted in Fig. \ref{fighr}, but severe consequences on 
the internal structure of the star. Fig. \ref{figtro}
 shows the predicted time dependence
of the central temperature-- density relation for selected values of $M_s$ and $n=2$. 
As expected, one finds that by increasing $M_s$
the efficiency of the extra-cooling  decreases.
  Above $M_s \simeq 5$ TeV/c$^2$  the effects on the 
evolutionary history of the stellar structure vanish. Even a quick inspection of
data in Fig. \ref{figtro} reveals that the assumption $M_s\sim 1.5$ TeV/c$^2$
(i.e. already above the current accelerator lower limit for $M_s$) is deeply affecting
 the structure so that one expects
strong observational consequences. As a matter of fact, by exploring the case 
$M_s$=1 TeV/c$^2$ (the previous lower limit) one finds that  RGB stars would fail to ignite 
Helium, running against the well-established evidence of Helium burning star in 
galactic GCs. Fig. \ref{figtro} shows that increasing the cooling for each given central 
density the central temperature is lower, as expected. According to well known 
prescriptions of  the stellar evolutionary theory, one can thus easily predict that 
the end of the RGB phase -- i.e. the central He ignition --  will be delayed and the mass of the 
He core at this stage will be larger.

To discuss this point in some detail, we show in Fig. \ref{figmc} the
He core mass at the central He ignition (RGB tip) for selected
assumptions about the value of $M_s$ and n=2.  As shown in the same
Fig. \ref{figmc} in order to cover the range of metallicity (Z)
spanned by the galactic GCs, computations have been performed for a
$0.8M_\odot$ star with Z=0.0002 and for $1M_\odot$ star with Z=0.02.

To constrain the value of $M_s$ one has now to discuss theoretical results in terms 
of observable quantities. In this context one finds that the extra-cooling is 
governing two main observational parameters: i) the luminosity 
of the RGB tip, ii) the luminosity of He burning HB stars. In both cases the  stronger the 
extra-cooling the larger is the predicted luminosity.

In this paper, we focus our attention on the first parameter only.
Several papers have already remarked the good agreement
between  observations and 
standard model theoretical predictions  \cite{DA90,SC98}. Such good agreement is shown 
also in Fig. \ref{figtip}
 where we report luminosity (in bolometric absolute magnitudes) 
of the brightest observed RGB star in clusters with different metallicities \cite{Fro83,Fer20}
as compared with theoretical predictions for the canonical scenario 
($M_s \rightarrow \infty$). According to the discussions given in several papers 
(see e.g. \cite{CDL,SC98}) the theoretical predictions should represent within 
about 0.1 mag. the {\em {upper envelope}} of the observed star luminosity, and this is
precisely 
what one finds in Fig. \ref{figtip}. However, the same figure shows that 
for finite value of $M_s$
 theoretical predictions move toward larger luminosity, in 
disagreement with observations. 

By inspection of data in Fig. \ref{figtip} one can conclude  conservatively that values of  
$Ms \leq$ 3 TeV/c$^2$ are definitively ruled out by the observational tests, whereas a 
lower limit of 4 TeV/c$^2$ appears reasonably acceptable. 
Thus this detailed evolutionary investigation has improved the crude 
estimate of ref \cite{Barger99} $M_s \gapprox 2 $ TeV/c$^2$.

\section{Concluding remarks}

We summarize here the main points of this paper:\\

\noindent 
i)Helioseismic constraints on the sound speed in the
 energy production region and on the photospheric helium
abundance rule out values of the $4+2$ dimensional 
Planck mass below $M_s= 0.3$ TeV/c$^2$.

\noindent
ii)The introduction of additional energy loss due to 
KK-graviton production cannot be a cure to the puzzles posed by 
SSM calculations. In particular, the predicted neutrino
signals would be even larger than those of the SSM.
  
\noindent
iii)Observational constraints
for Red Giant stars evolving in galactic globulars imply 
$M_s > 3-4$ TeV/c$^2$. This bound is stronger than
that provided by accelerator, thus 
indicating how useful it is, and hopefully it will be, the synergetic use
of terrestrial and stellar laboratories.

\acknowledgments

We are extremely grateful to Z. Berezhiani, D. Comelli and F. Villante
for discussions. This work is co-financed by the
 Ministero dell'Universit\`a e della Ricerca Scientifica e
Tecnologica (MURST) within the ``Astroparticle Physics'' project.

\appendix

\section{Star energy-loss via KK-gravitons}

The energy loss rate per unit mass  
due to escaping KK gravitons has been calculated in
\cite{Barger99}. Three  processes are important
for KK-graviton production
in the sun and in the red giants.
The relevant formulas 
are collected below, in natural units as well as in units
more useful for implementation in a stellar evolution code.
For a comparison, the energy production rate per unit mass
due to the pp-chain is also parametrized.

\subsubsection{Photon photon annihilation to KK gravitons: 
$\gamma+\gamma \rightarrow grav$}

When $n$ extra dimensions are effective, the Newtonian
interaction potential [ $ V \sim 1/(M_{pl}^2 r) $]is modified to
$V \sim  1/(M_s^{n+2} r^{1+n})$, so that the coupling of each particle
to the gravitational field is proportional to $ 1/M_s^{1+n/2}$
and KK-graviton 
production cross sections are proportional to the square
of this quantity. 
A thermal photon gas is uniquely specified
by its temperature T and fundamental physical constants
$(\hbar, c$ and $K_B$) so that dimensional considerations
fix the dependence of the energy loss rate per unit volume $Q_{\gamma}$.
In natural units, this has dimension of [Energy]$^5$, so that
one has $ Q_{\gamma}= A_{\gamma}(n) M_s^{-n-2}T^{n+7}$ where
$A_{\gamma}(n)$ are numerical coefficients given in eq. (7) of 
\cite{Barger99}. The energy loss rate per unit mass is obtained by
dividing Q by the mass density $\rho$. When temperature
is in expressed in Kelvin degress, density in g/cm$^3$ and
$M_s$ in TeV/c$^2$, the energy loss $\epsilon_{\gamma}$ in erg/g/s
is thus:
\begin{equation}
\label{eqeg1}
n=2 \quad \epsilon_{\gamma}= 7.25 \, \cdot 10^{-66} \frac{T^9} {\rho M_s^4}
\end{equation}
\begin{equation}
\label{eqeg2}
n=3 \quad \epsilon_{\gamma}= 4.42 \, \cdot 10^{-82} \frac{T^{10}} {\rho M_s^5}
\end{equation}

\subsubsection{Gravi-compton Primakoff scattering:
$\gamma +e \rightarrow e+grav$}

The expression for the energy loss is in this case:
\begin{equation}
\epsilon_{GCP}= B(n) \frac{\alpha}{m_e} \frac{n_e}{\rho}
                 \frac{T^{n+5}} {M_s^{n+2}}
\end{equation}
where the numerical coefficients $B(n)$ are found in eq. 15
of \cite{Barger99}, $m_e (n_e)$ is the
electron mass (numerical density). The dependence on 
$M_s$ is easily understood from the previous considerations,
$\alpha$ comes in  from 
electro-magnetic coupling of the electron
and the factor $n_e/\rho$ clearly expresses the proportionality
to the electron number per unit mass. 
Dimensional analysis
is not sufficient to fully specify the dependence on
temperature due to the presence of another
mass scale, $m_e$, which is relevant  
for non-relativistic electrons, see \cite{Arkani98b}.
In the same units as in eq. (\ref{eqeg1},\ref{eqeg2}):
\begin{equation}
\label{eqgcp1}
n=2  \quad \epsilon_{GCP}= 1.69 \, \cdot 10^{-78} \frac{T^7 n_e} {\rho M_s^4}
\end{equation}
\begin{equation}
\label{eqgcp2}
n=3  \quad \epsilon_{GCP}= 5.60 \, \cdot 10^{-94} \frac{T^{8}n_e} {\rho M_s^5}
\end{equation}

\subsubsection{Gravi-Bremsstrahlung: $e+Z \rightarrow e+Z+grav$}

The energy loss is now:
\begin{equation}
\epsilon_{GB}= C(n) \alpha^2 \frac{n_e}{\rho}
                 \frac{T^{n+1}} {M_s^{n+2}} \Sigma_j n_j Z_j^2
\end{equation}
where $n_j$ is the number density of nuclei with atomic
number $Z_j$ and the numerical factors  $C(n)$
are given in eq. (21) of \cite{Barger99}.
In the same units of eqs. (\ref{eqeg1},\ref{eqeg2}) one has:
\begin{equation}
\label{eqgb1}
n=2\,: \quad  \epsilon_{GB}= 5.86 \, \cdot 10^{-75} \frac{T^3 n_e} {\rho M_s^4}
\Sigma_j n_j Z_j^2 
\end{equation}
\begin{equation}
\label{eqgb2}
n=3\,:  \quad \epsilon_{GB}= 9.74 \, \cdot 10^{-91} \frac{T^{4}n_e} {\rho M_s^5}
\Sigma_j n_j Z_j^2 
\end{equation}

~\\

The total energy loss due to KK-graviton production is
\begin{equation}
\epsilon_{KK}=\epsilon_{\gamma} + \epsilon_{GCP} + \epsilon_{GB}
\end{equation}

~\\

It is useful to compare the above energy losses with the 
e.m. energy production rate per unit mass  from the pp-chain.
The slowest reaction of the chain is the $p+p\rightarrow d+e^+ +\nu_e$
which in the temperature region of interest for the sun, has a rate 
$<\sigma v>_{pp}= A T^{3.83}$, with $A=4.398 \,\cdot 10^{-71}$ cm$^3$/s and $T$
is expressed in Kelvin.
The energy production rate  per unit mass  through the ppI termination of the
pp-chain is:
\begin{equation}
\label{epsipp}
\epsilon_{ppI}= \frac{1}{4} \rho \frac{X^2}{m_H^2} Q_{em} <\sigma v>_{pp}
\end{equation}
where $Q_{em}=26.1$ MeV is the average e.m. energy released in the 
$4p+2e^- \rightarrow ^4He+2\nu_e$, $X $ is the H-mass fraction
and $m_H$ is the hydrogen mass. In the same units as in Eq. 
(\ref{eqeg1}) one has:
\begin{equation}
\epsilon_{ppI}= 2.97 \, \cdot 10^{-27} X^2 \rho T^{3.83}
\end{equation}
As well known the ppI  branch is the main energy source in the Sun. 
Everywhere it gives the strongest contribution to the energy production
rate: 
\begin{equation}
\label{epsinuc}
\epsilon_{nuc}= \epsilon_{ppI}+\epsilon_{ppII}+\epsilon_{ppIII}+\epsilon_{CNO} \, .
\end{equation}


\begin{table}
\caption{  Physical and chemical 
properties of the solar center, according to the
SSM of [16].}
\begin{tabular}{lc}
$T$[K]		&  $1.569 \,\cdot 10^7$\\
$\rho$ [g/cm$^3$]	& 152\\
$X$		& 0.33867\\
$Y$		& 0.64014\\
$\Sigma_j \frac{X_j}{A_j} Z_j^2 $ & 1.06
\end{tabular}
\label{tabssm}
\end{table}
 
\begin{table}
\caption{KK-energy loss and production
rates calculated at the solar center (see Appendix for definitions).
Rates are in erg/g/s and the
 $4+n$ dimensional Planck mass $M_s$ is in TeV/c$^2$.}
\begin{tabular}{lcc}
		       &   $n=2$	   &       $n=3$  \\
$\epsilon_\gamma$    & $2.75\,\cdot 10^{-3}M_s^{-4}$  & $2.63\, \cdot 10^{-12}M_s^{-5}$\\
$\epsilon_{GB}$      & $1.59\, \cdot 10^{-4}M_s^{-4}$  & $8.26\, \cdot 10^{-13}M_s^{-5}$\\
$\epsilon_{GCP}$     & $8.82\, \cdot 10^{-4}M_s^{-4}$  & $2.29\, \cdot 10^{-12}M_s^{-5}$\\
$\epsilon_{KK}$      & $3.79\, \cdot 10^{-3}M_s^{-4}$  & $5.75\, \cdot 10^{-12}M_s^{-5}$\\
\hline
$\epsilon_{ppI}$       & $10.5$ & $10.5$ 
\end{tabular}
\label{tabepsi}
\end{table}

\begin{table}
\caption{Fractional differences, (model-SSM)/SSM, between the calculated 
properties  of a solar model with $M_s=0.2$ TeV/c$^2$ ($n=2$)
and the SSM.}
\begin{tabular}{lc}
initial composition & \\
$Y_{in}$  & -2.3\%\\
$Z_{in}$  & +0.88\%\\
\hline
convective envelope & \\
$Y_{ph}$  &-2.5\%\\
$R_{b}$   &+0.08\%\\
$T_b$     &-0.72\%\\
\hline
solar center & \\
$X_c$       &-5.6\%\\
$Y_c$       &+3.0\%\\
$T_c$       &+1.7\%\\
\hline
neutrino fluxes & \\
$Be$       & +24\%\\
$B $       & +52\%
\end{tabular}
\label{tab02}
\end{table}

\begin{table}
\caption{ The best fit parameters for Eq. \ref{eqfit}.}
\begin{tabular}{lcc}
          & $m$[GeV/c$^2$] & $\alpha$\\
$u_{0.2}$ & 60    & 4.3 \\
$Y_{ph}$  &  80   & 4.1 \\
\end{tabular}
\label{tabalfa}
\end{table}


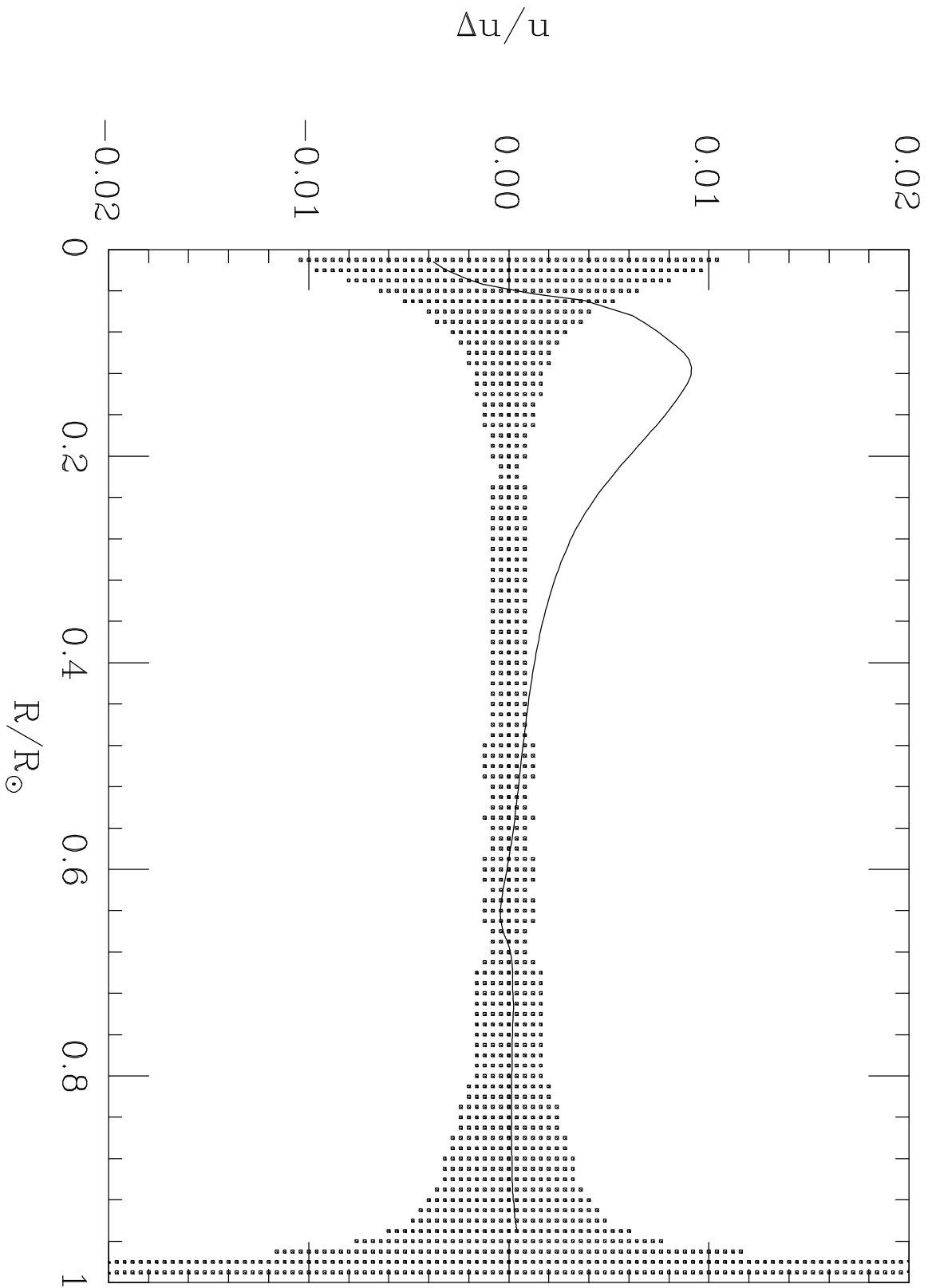
\begin{figure}
\caption{Fractional variation with respect to the SSM prediction,
(model-SSM)/SSM, of the squared 
isothermal sound speed $u(r)=P/\rho$ in the solar model
 with $M_s=0.2$ TeV/c$^2$ and $n=2$. The shaded area corresponds to the 
``$1 \sigma$'' or statistical helioseismic uncertainty, see [6].}
\label{fig02}
\end{figure}

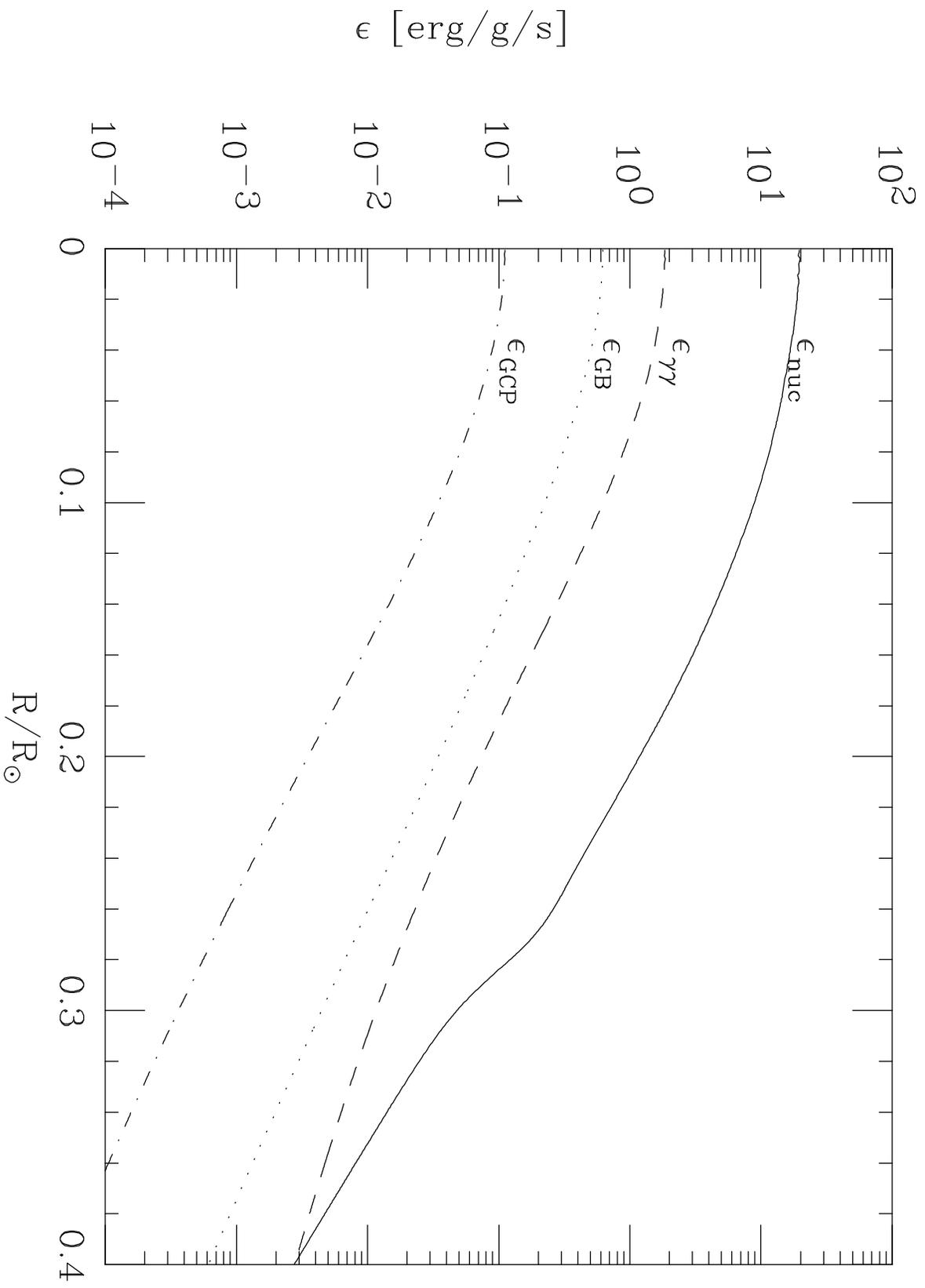
\begin{figure}
\caption{Energy losses due to  KK-graviton production along the
solar structure of the model
with $M_s=0.2$ TeV/c$^2$ and $n=2$.
 Dashed line corresponds to the photon-photon annihilation,
dash-dotted line to gravi-Compton-Primakoff effects, dotted line to 
gravi-bremsstrahlung process. For a comparison also the nuclear
energy production (full line) is shown.}
\label{figepsi}
\end{figure}

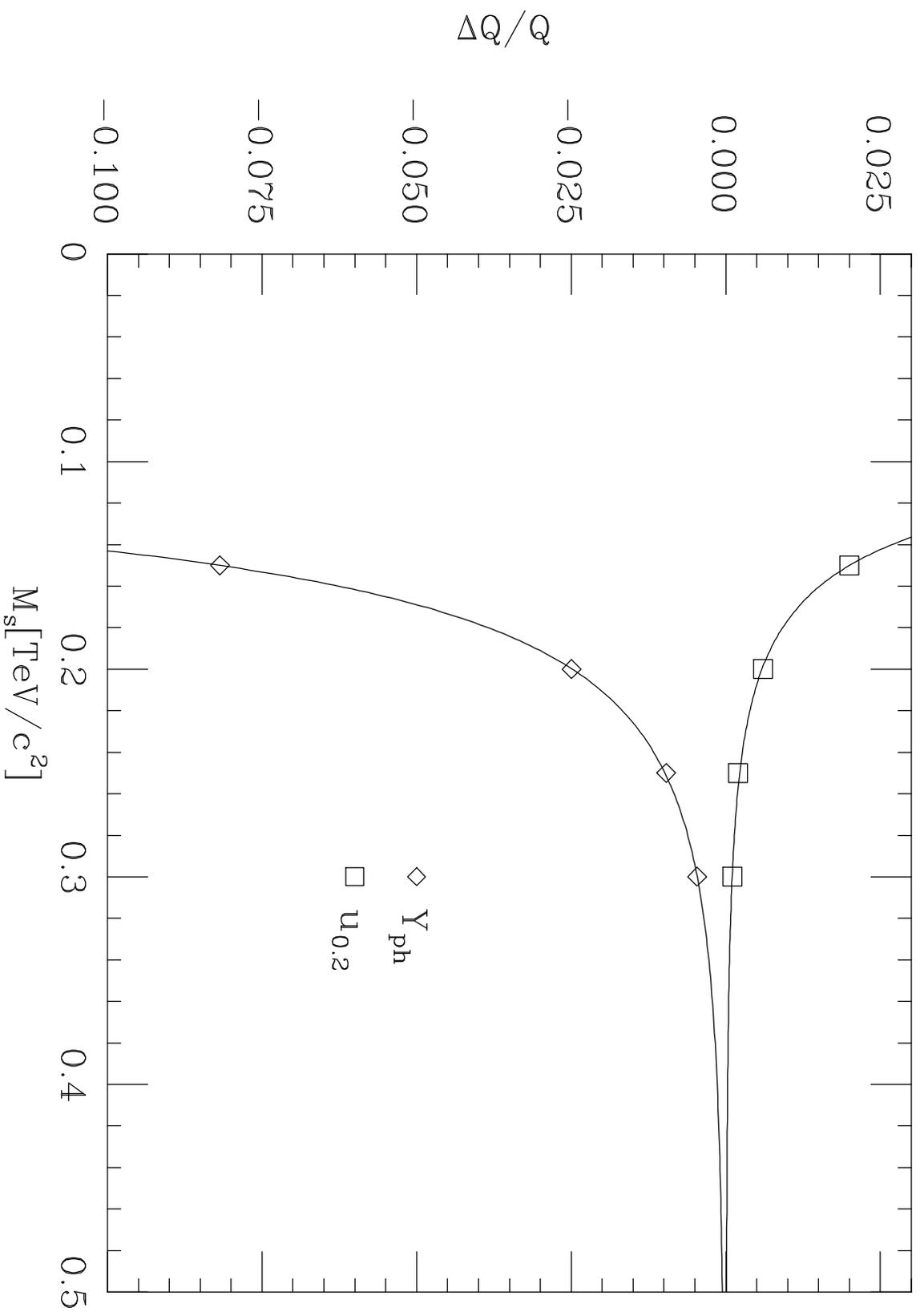
\begin{figure}
\caption{Fractional variation with respect to the SSM predictions of the
quantities $Y_{ph}$ (diamonds) and $u_{0.2}$ (squares) in the solar models 
 with  different values of $M_s$.}
\label{figdelta}
\end{figure}

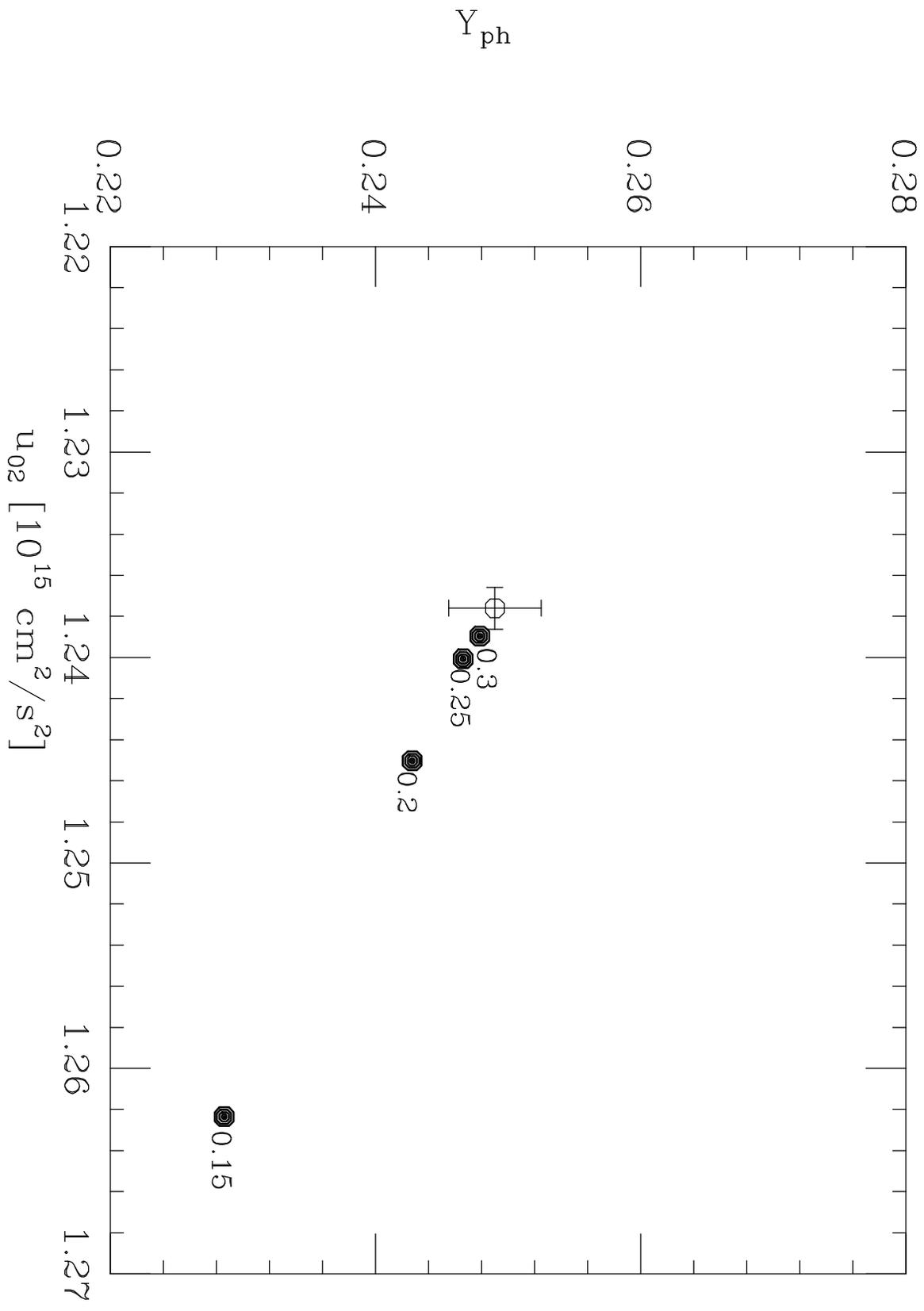
\begin{figure}
\caption{The photospheric helium abundance $Y_{ph}$ and the value of 
$u=P/\rho$ at $R=0.2 R_\odot$: a) as constrained by helioseismology, see
Eqs. (\ref{eqyph},\ref{equ02}); b) as modified
by models with the indicated value of $M_s$, in TeV/c$^2$).}
\label{figpiano}
\end{figure}

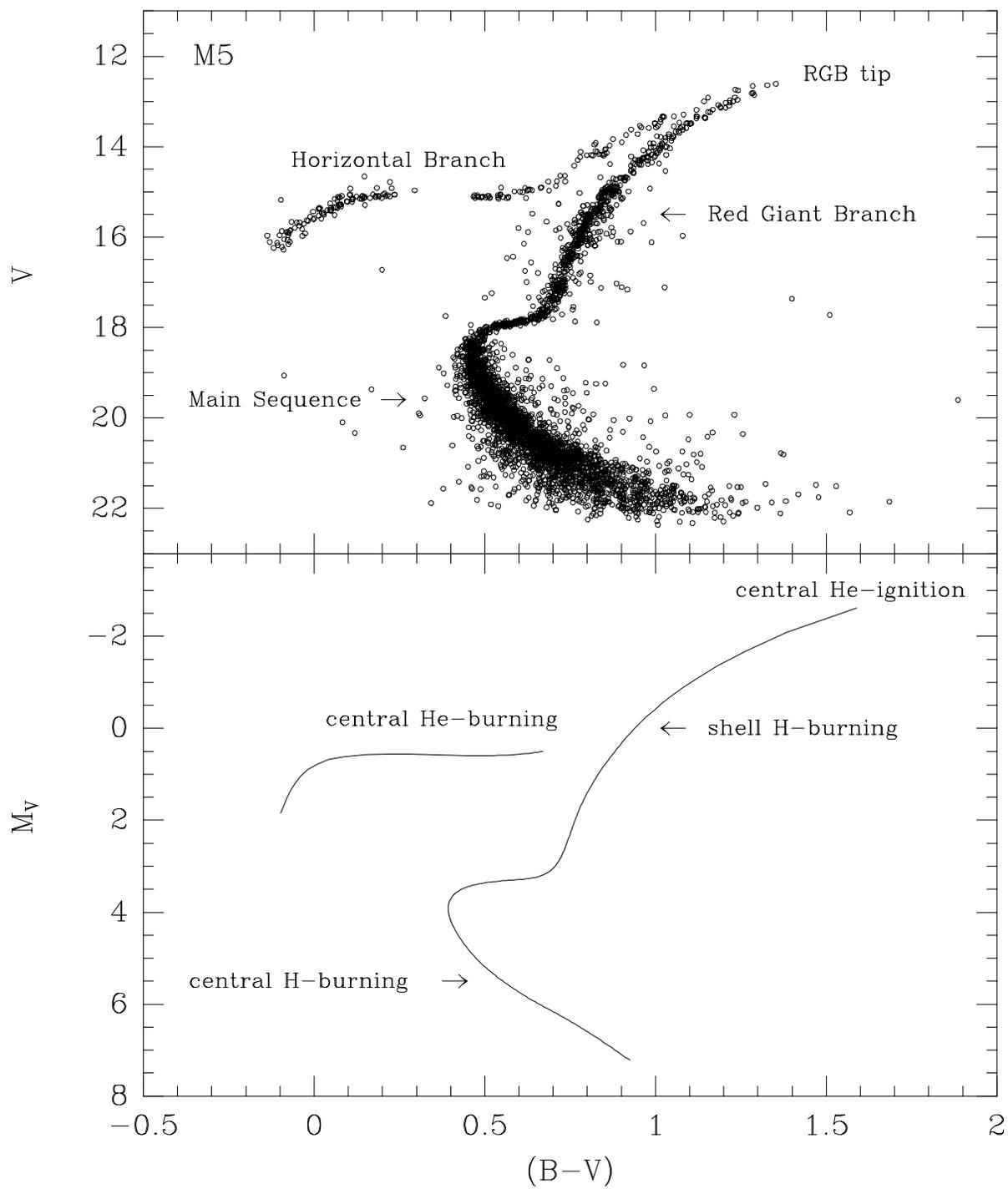
\begin{figure}
\caption {Upper panel: typical observational Hertzprung-Russel diagram
for a galactic GC. Lower panel: the corresponding theoretical
Hertzprung-Russel diagram.  The most relevant evolutionary phases are
shown.}
\label{fighr}
\end{figure} 

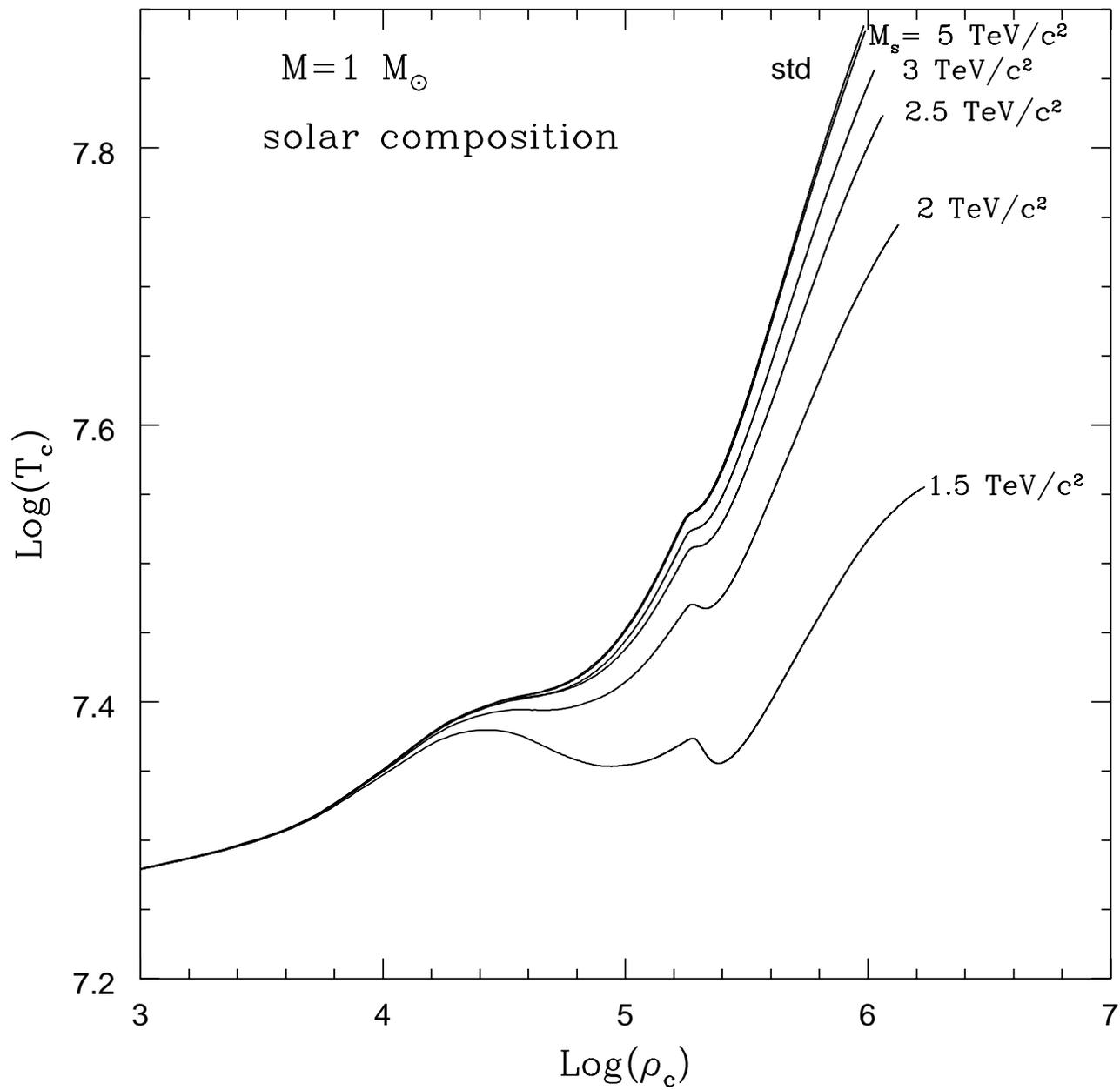
\begin{figure}
\caption{Time behaviour
of the central temperature -- density relation for a solar model from the Main sequence
to the ignition of central He burning as predicted by present evolutionary calculations
for the canonical case (std) and for the labelled values of $M_s$ and $n=2$.}
\label{figtro}
\end{figure}

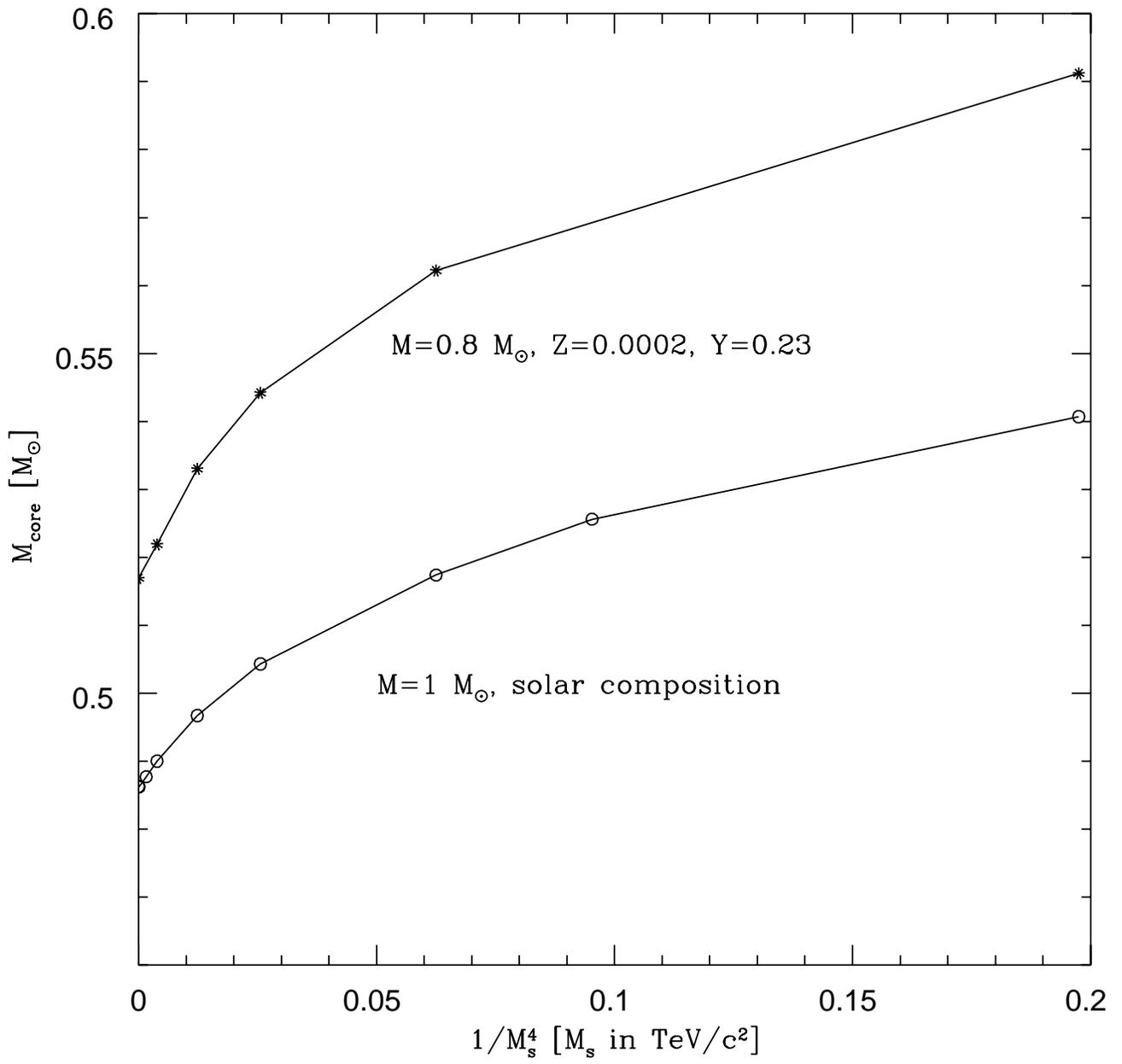
\begin{figure}
\caption{The mass, in solar units, of the He core at the central He ignition (RGB tip)
 as a function of $M_s$ for n=2.  In order to cover the range of metallicity (Z)
spanned by the galactic GCs, computations have been performed for a
$0.8M_\odot$ star with Z=0.0002 and for $1M_\odot$ star with Z=0.02.}
\label{figmc}
\end{figure}

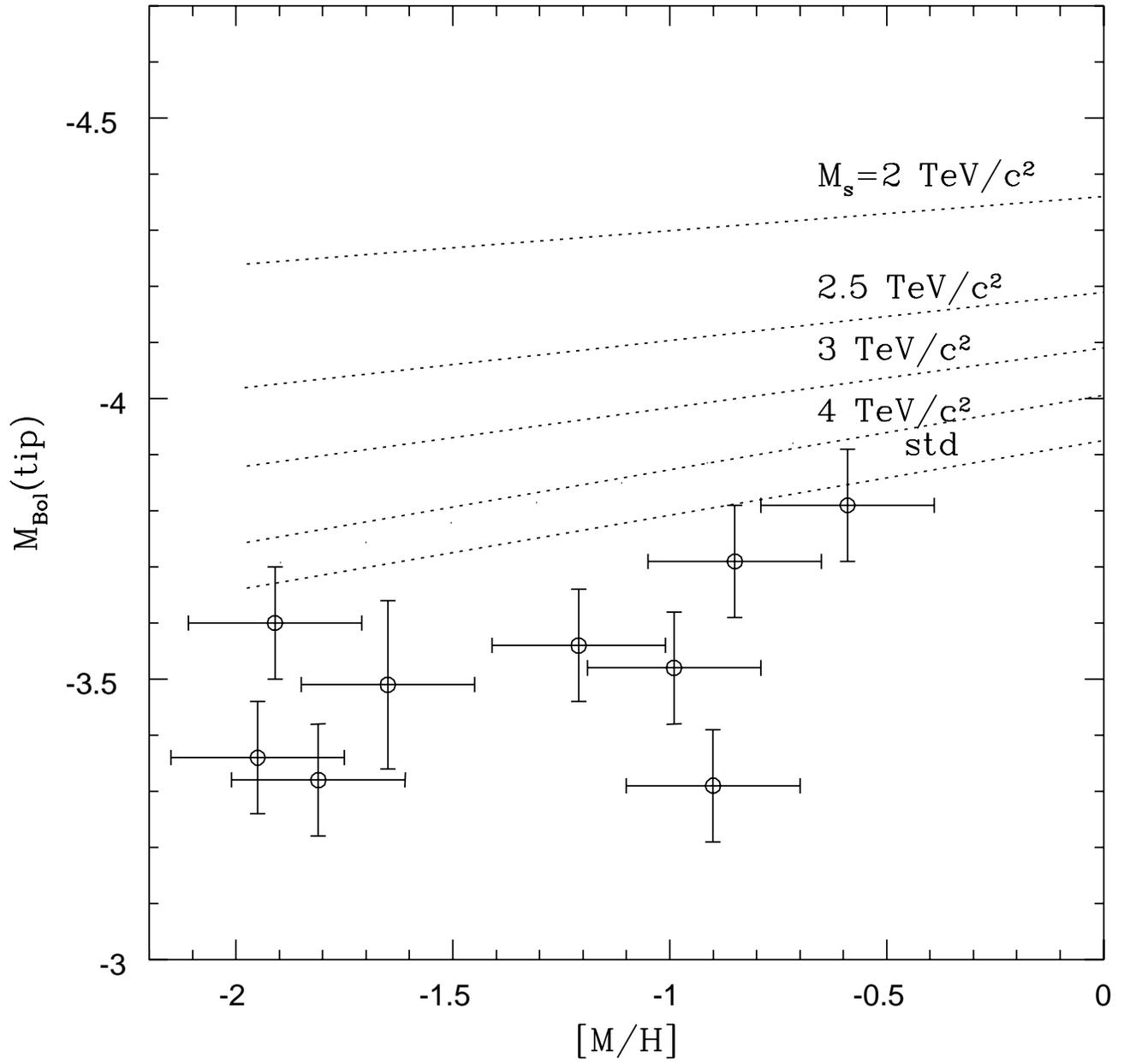
\begin{figure}
\caption{Luminosity (in bolometric absolute magnitudes) 
of the brightest observed RGB star (RGB tip) in clusters with different metallicities
 ([M/H] = Log(M/H)$_{\rm star}$ - Log(M/H)$_{\odot}$, where M is fractional abundance
by mass of all the elements heavier than Helium) [24,25]
 as compared with theoretical predictions for the canonical scenario (std)
and for models with energy losses due to KK-gravitons with the labelled values
of $M_s$ and $n=2$.}
\label{figtip}
\end{figure}

\end{document}